\documentclass[aps,pra,groupedaddress,showpacs,preprint]{revtex4-1}
\usepackage[english]{babel}
\usepackage{graphicx}

\begin{document}
\title{Potential splitting approach to e-H and e-He${}^+$ scattering with zero total angular momentum} 

\author{E.~Yarevsky$^1$}
\email[E-mail: ]{e.yarevsky@spbu.ru}
\author{S.~L.~Yakovlev$^1$}
\email[E-mail: ]{s.yakovlev@spbu.ru}
\author{N.~Elander$^2$}
\email[E-mail: ]{elander@fysik.su.se}

\affiliation{$^1$ St Petersburg State University, 7/9 Universitetskaya nab., St. Petersburg, 199034, Russia}
\affiliation{$^2$ Chemical Physics Division, Department of Physics, AlbaNova University Centre, Stockholm University, 106 91 Stockholm, Sweden, EU}

\begin{abstract}
An approach based on splitting the reaction potential into a finite range part and a long range tail part to describe few-body scattering in the case of a Coulombic interaction is proposed. 
The solution to the Schr\"odinger equation for the long range tail of the reaction potential is used as an incoming wave. 
This reformulation of the scattering problem into an inhomogeneous Schr\"odinger equation with asymptotic outgoing waves makes it suitable for solving with the exterior complex scaling technique.
The validity of the approach is analyzed from a formal point of view and demonstrated numerically, where the calculations are performed with the finite element method. 
The method of splitting the potential in this way is illustrated with calculations of the electron scattering on the hydrogen atom and the positive helium ion in energy regions where resonances appear.
\end{abstract}


\maketitle
%


\section{Introduction} \label{intro}
Difficult fundamental problems should, if possible, be addressed to few-body physics which offers detailed, numerically almost exact, solutions. 
Electron scattering off the hydrogen atom and the helium cation are just such problems, and being of fundamental importance to atomic physics, any developed approach is worth testing with these problems. 
By comparing very detailed theoretical and computational results on the one hand to experimental results on the other, one can obtain guidelines to the development of an understanding of more complicated systems. 
The simplest many-body electron-scattering problem is no doubt electron scattering on the hydrogen atom or a hydrogen like ion. 
These problems can be treated by few-body techniques as well as, for example, advanced many-body methods.

The complicated boundary conditions at large distances are a major difficulty for the present kind of problems, especially when the long-range
Coulomb interaction is present~\cite{FadMerk}.  
To date, several methods have been developed for constructing solutions to the three-body scattering problem (see~\cite{PhysRep.520.135} and references therein). 
Considerable efforts to avoid using the explicit form of the asymptotic nature of the wave function have been made, where several of these are based on complex scaling theory~\cite{CS}.
A modification of the method of splitting the potential into two, based on the introduction of a cut-off of the reaction potential at some distance $R$, was successfully applied to collision systems with a long-range, non-Coulomb, potential~\cite{Rescigno1997}.
As a cut-off potential is not an analytic function, exterior complex scaling (ECS)~\cite{r:ecs1, r:ecs2} beyond the point $R$  has been employed. 
However, this modified approach cannot be applied directly to the Coulomb scattering problem since a cut-off of the Coulomb potential at any finite distance essentially distorts the asymptotic behavior of the solution at large separations~\cite{FadMerk}.
Other ways of employing complex scaling to scattering problems have appeared in Ref.~\cite{Gasaneo2012, Zaytsev2016} where complex rotation of the basis functions rather than of the system Hamiltonian has been proposed and studied.

In several recent studies, we have reported a method which accurately solves the Schr\"odinger equation for the Coulomb scattering problem using ECS ~\cite{PRA-2011, EPL-2015, JPB-2015}.
This particular approach is based on the sharp splitting of the Coulomb potential to construct the distorted incident wave which is generated by the asymptotic tail of the Coulomb potential. 
This reformulation of the Coulomb scattering problem makes it suitable to the application of the ECS technique. 
The present study is the next step in the development of a theory and the necessary computational tools to be able to calculate three-body many-channel scattering for systems which are described with numerical potentials having a known asymptotic analytical form.
For example, the reaction H$^+$ + H${}_2^-$ $\to$ H${}_3$ $\to$ H + H${}_2$ is just such a reaction. 
The method is an extension of our previously presented resonance theory and code~\cite{r:eyrescode}.

The paper is organized as follows. 
In Section~\ref{theory}, the formalism of the potential splitting approach is developed, while in Section~\ref{results} our computational method and the results for the electron-hydrogen and electron-He${}^+$ scattering processes are discussed. 
Atomic units are used throughout the paper.

\section{Theoretical approach} \label{theory}

For the example of electron scattering  off a hydrogen-like atom, the Hamiltonian for the full problem is written in terms of the electron-nucleus distances $\mathbf{r}_1$, $\mathbf{r}_2$ as 
\begin{equation} \label{Hamilt_full}
H = - \frac{1}{2} \Delta_{\mathbf{r}_1}
    - \frac{1}{2} \Delta_{\mathbf{r}_2}
    - \frac{Z}{|\mathbf{r}_1|} - \frac{Z}{|\mathbf{r}_2|}
    + \frac{1}{|\mathbf{r}_1-\mathbf{r}_2|},
\end{equation}
where $Z$ is the nuclear charge, and the nucleus mass is assumed to be infinite. 
As the total angular momentum is conserved, processes with different angular momenta can be studied independently. 
In the present paper, we only consider the scattering process with a zero total angular momentum. 
The reason for this restriction at this stage of our work is our desire to avoid unnecessary technical complications and allow us to focus on the approach and its validity. 
The introduction of arbitrary total angular momenta is ongoing but remains as our next step in the development process. 
The projection $H_0$ of the full Hamiltonian~(\ref{Hamilt_full}) on the subspace of zero total angular momentum can be written as 
\begin{equation} \label{Hamilt}
H_0 = {H}^K + V(r_1,r_2,\theta).
\end{equation}
Here the kinetic energy is given by
\begin{equation}
H^K = - \frac{1}{2} \frac{\partial^2}{\partial r_1^2} - \left(\frac{1}{2 r_1^2}+\frac{1}{2 r_2^2} \right)
\left(
\frac{\partial^2}{\partial {\theta}^2} + \cot{\theta} \frac{\partial}{\partial {\theta}} 
\right)
-\frac{1}{2} \frac{\partial^2}{\partial r_2^2} ,
\end{equation}
where $r_i=|\mathbf{r}_i|$, i=1,2, and $\theta$ is the angle between the vectors $\mathbf{r}_1$ and $\mathbf{r}_2$.
The total potential, $V(r_1,r_2,\theta)$, is the sum of the Coulomb pair-wise potentials:
\begin{equation} \label{full_pot}
V(r_1,r_2,\theta) = -\frac{Z}{r_1} -\frac{Z}{r_2} +V_{12}(r_1,r_2,\theta),
\end{equation}
where the electron-electron interaction $V_{12}(r_1,r_2,\theta) =1/|\mathbf{r}_1-\mathbf{r}_2|= 1/\sqrt{r_1^2+r_2^2-2r_1 r_2 \cos\theta}$.


The solution of the Schr\"odinger equation $(H_0-E)\Psi(r_1,r_2,\theta)=0$ must both satisfy the boundary conditions  $\Psi(r_1,0,\theta)= \Psi(0,r_2,\theta)=0$, and have the correct asymptotic behaviour at large distances. The latter requirement will be discussed  below.

As the electrons are identical fermions, the proper symmetry of the wave function with respect to the permutation of the electron coordinates is required.
The symmetrized wave function $\Psi^S$ is defined as $\Psi^S = P^S \Psi$, where the symmetrization operator $P^S$ is given by the standard expression
\begin{equation} \label{P_sym}
P^S= \frac{1}{\sqrt{2}} (1+(-1)^S P_{12}) .
\end{equation}
Here $S=0,1$ stands for singlet or triplet scattering, respectively.
The permutation operator $P_{12}$ interchanges the electrons 1 and 2.
As the permutation operator commutes with the Hamiltonian $H_0$, the symmetrized wave function obeys the same Schr\"odinger equation $(H_0-E)\Psi^S = 0$.
For the sake of clarity, our derivations will be made for the function $\Psi$, and the symmetrization will be done at the final stage of the formalism by applying the operator $P^S$.


\subsection{Potential splitting approach}
The solution of the scattering problem for the Schr\"odinger equation with the Hamiltonian~(\ref{Hamilt}) involves complicated boundary conditions~\cite{FadMerk}.
They are hard to implement especially for the Coulomb interactions.
Here we describe the potential splitting approach~\cite{PRA-2011, EPL-2015, JPB-2015} which allows us to solve the Coulomb scattering problem without explicit use of the asymptotic form of the wave function. 
For definiteness we imply that electron 1 collides with the bounded complex of electron 2 and the nucleus.

Let $\chi^R(r)$ be the indicator of the domain $r \ge R$, i.e.
\begin{equation}
\chi^R(r) = \left\{
\begin{array}{ll}
0, & r < R\\
1, & r \ge R ,
\end{array}
\right.
\end{equation}
and $\chi_R = 1-\chi^R$ be its complementary partner.

The reaction potential $V^{\rm reac}(r_1,r_2,\theta)$
\begin{eqnarray}
V^{\rm reac}(r_1,r_2,\theta) 
 &=& V(r_1,r_2,\theta) - \left( -\frac{Z}{r_2}\right) \nonumber \\
 &=& -\frac{Z}{r_1} +V_{12}(r_1,r_2,\theta)
\end{eqnarray}
is split  into the sum of the core $V_R$ and the tail $V^R$ components
\begin{equation}
V^{\rm reac}(r_1,r_2,\theta)  = V_R(r_1,r_2,\theta)  + V^R(r_1,r_2,\theta) ,
\end{equation}
where
\begin{equation}
V_R  =  V^{\rm reac} \chi_R(r_1), \quad
V^R  =  V^{\rm reac} \chi^R(r_1).
\end{equation}
The distorted incident wave $\Psi^R(r_1,r_2,\theta)$ is the solution to the scattering problem with the sum of the bound pair potential $-Z/r_2$ and the tail potential $V^R(r_1,r_2,\theta)$:
\begin{equation} \label{eq_psiR}
\left[H^K -\frac{Z}{r_2} + V^R(r_1,r_2,\theta) - E\right] \Psi^R(r_1,r_2,\theta) = 0.
\end{equation}

The function $\Phi(r_1,r_2,\theta)$ now is defined as the difference $\Phi \equiv\Psi - \Psi^R$, and it satisfies the driven Schr\"{o}dinger equation
\begin{equation} \label{driven}
\left( H_0 - E \right) \Phi = - V_R \Psi^R.
\end{equation}
This constitutes the main equation of the potential splitting approach. 
Its right hand side (r.h.s.) is of finite range with respect to the variable $r_1$. 
Furthermore, $\Phi$ behaves as a superposition of pure outgoing waves in all asymptotic regions.
Therefore, Eq.~(\ref{driven}) is suitable for ECS~\cite{r:ecs1, r:ecs2, r:ecs3} with the exterior scaling radius $Q \ge R$. 
After ECS, the function $\Phi$ becomes an exponentially decreasing function, implying that boundary conditions equal to zero can be used in order to solve the ECS-transformed equation~(\ref{driven}).

Let us now discuss the construction of the distorted incident wave $\Psi^R$. 
First, we determine the asymptotic initial state by introducing the function $\Psi^R_0(r_1,r_2,\theta)$ as the solution to Eq.~(\ref{eq_psiR}), where the potential $V^R(r_1,r_2,\theta)$ is replaced by its leading term in the incident configuration
\begin{equation}
V^R_{\cal C}(r_1)= - \frac{(Z-1)}{r_1} \chi^R(r_1) .
\end{equation}
The variables in this case can be separated and the solution, in the case of zero total angular momentum, can be explicitly derived~\cite{EPL-2015, ourJPhysA2010} as
\begin{eqnarray} \label{psi0_0}
& \Psi^R_0 &=
\frac{i^{\ell_i}}{p_i r_1} \varphi_{n_i,\ell_i}(r_2) Y_{\ell_i,0}(\theta,0) \nonumber \\
&\times& \left\{
\begin{array}{ll}
a^R_{\ell_i} {\hat j}_{\ell_i}(p_{i} r_1), & r_1 < R. \\
e^{i\sigma_{\ell_i}} F_{\ell_i}(\eta_i, p_i r_1)
+ {\cal A}^R_{\ell_i} u^+_{\ell_i}(\eta_i, p_i r_1), & r_1 \geq R.
\end{array}
\right.
\end{eqnarray}
Here the direction of the $z$-axis is chosen to coincide with the direction of the incident momentum $\mathbf{p}_{i}$.
The function $\tilde{\varphi}_i (\mathbf{r}_2) = r_2^{-1} \varphi_{n_i,\ell_i}(r_2) Y_{\ell_i,0}(\theta,0)$ is the target bound state wave function for the two-body system with the energy $\varepsilon_i$ and quantum numbers $n_i$, $\ell_i$.
The value of the incident momentum $p_i$ is related to the total scattering energy, $E$, as $E=p_i^2/2 + \varepsilon_i$.
${\hat j}_\ell$ is the Riccati-Bessel function, the Sommerfeld parameter is given by  $\eta_i=-(Z-1)/p_i$.
The Coulomb outgoing wave function 
\[
u^+_{\ell}(\eta_i, p_i r_1) = e^{-i\sigma_\ell} (G_\ell(\eta_i, p_i r_1) + i F_\ell(\eta_i, p_i r_1))
\]
is expressed in terms of the regular $F_\ell$ and irregular $G_\ell$ Coulomb wave functions~\cite{AbrSt} and the Coulomb phase shift $\sigma_{\ell}$.
The coefficients $a^R_\ell$ and ${\cal A}^R_\ell$ are defined as~\cite{ourJPhysA2010}
\begin{eqnarray}
a^R_\ell & = & e^{i\sigma_{\ell}}W_R(F_\ell,u^+_\ell)/W_R({\hat j}_\ell, u^+_\ell), \nonumber \\
{\cal A}^R_\ell & = & e^{i\sigma_{\ell}}W_R(F_\ell,{\hat j}_\ell)/W_R({\hat j}_\ell, u^+_\ell), \label{amplA^R_l}
\end{eqnarray}
where the Wronskian, $W_R(f,g)=fg' - f'g$, is calculated at $r = R$.

The function $\Psi^R_0$~(\ref{psi0_0}) does not satisfy Eq.~(\ref{eq_psiR}) exactly, $\Psi^R \neq \Psi^R_0$.
The full solution to Eq.~(\ref{eq_psiR}) can then be represented as 
\begin{equation}\label{Psi^R_10}
\Psi^R = \Psi^R_0 + \Psi^R_1
\end{equation} 
for which the function $\Psi^R_1$ satisfies the inhomogeneous equation
\begin{equation}\label{eq_psiR1}
\left(H^K - \frac{Z}{r_2} + V^R - E\right)\Psi^R_1 = -(V^R-V^R_{\cal C}) \Psi^R_0 .
\end{equation}
All incoming waves in $\Psi^R$ are included in the function $\Psi^R_0$, implying $\Psi^R_1$ contains outgoing waves only. 
Therefore, $\Psi^R_1$ remains bounded after the ECS transformation and approaches zero at large distances.
Equation~(\ref{eq_psiR1}) is thus of the type which can be solved by the ECS approach. 
Indeed, in the region where $\tilde{\varphi}_i(\mathbf{r}_2)$ is not negligible (i.e. $r_2 \le \rm{const}$), one obtains 
\begin{equation} \label{correction_estim}
V^R(r_1,r_2,\theta)-V^R_{\cal C}(r_1) \sim O\left(r^{-2}_1 \right)
\end{equation}
as $r_1 \to \infty$. 
The non-Coulomb asymptotic tail of such a potential can be truncated at some $r_1=R'$, $R' > R$, and the ECS approach with the exterior scaling radius $Q' \ge R'$ can be applied as shown in~\cite{CBR2004}.

In the same way as in Eq.~(\ref{Psi^R_10}), the solution $\Phi$ of the full problem~(\ref{driven}) can be represented as
\begin{equation} \label{Phi_10}
\Phi = \Phi_0 + \Phi_1,
\end{equation} 
where the functions $\Phi_0$, $\Phi_1$ are the solutions to the equations
\begin{eqnarray} \label{Phi_i_eq}
\left(H^K - \frac{Z}{r_2} + V^{\rm reac} - E\right) \Phi_i &=& 
\left(H_0 - E\right) \Phi_i \nonumber \\
&=& - V_R \Psi^R_i, \quad i=0,1.
\end{eqnarray}
Thus the total wave function $\Psi$ is given by
\begin{equation} \label{psi_sum}
\Psi = \Psi^R_0 + \Phi_0 + \Psi^R_1 + \Phi_1.
\end{equation} 
The two last terms in this equation vanish when $R \to \infty$.
However, for moderate values of $R$, their contributions might not be negligible.
We discuss their influence in the next section.

In order to completely restore the function $\Psi$, we should first solve Eq.~(\ref{eq_psiR1}) for $\Psi^R_1$.
Here, there exist two possibilities.
We can construct $\Psi^R=\Psi^R_0+\Psi^R_1$, and solve Eq.~(\ref{driven}) for $\Phi$.
Alternatively, we can solve the two equations~(\ref{Phi_i_eq}) for $\Phi_0$ and $\Phi_1$, and determine the function $\Psi$ with Eq.~(\ref{psi_sum}).
The second approach demands more computational effort. 
On the other hand, it allows each of four contributions in Eq.~(\ref{psi_sum}) to be determined independently, so one can analyze and compare corresponding amplitudes. 
It is for this reason why we adopt the second method in the present paper.

As mentioned earlier, the wave function for the system should be properly symmetrized with respect to the permutation of electrons.
After applying the symmetrization operator $P^S$ to Eq.~(\ref{Phi_i_eq}), the symmetrized solutions $\Phi^S_i = P^S \Phi_i$ are given by:
\begin{equation} \label{Phi_i_eq_S}
\left(H_0 - E\right) \Phi^S_i = - P^S V_R \Psi^R_i, \quad i=0,1.
\end{equation}

\subsection{Asymptotic behaviour of the scattering wave function}

When solving Eq.~(\ref{Phi_i_eq_S}) with ECS, we obtain the wave function $\Phi^S$ in the region $r_1, r_2 \le Q$.
The next step is to calculate the amplitudes and cross sections corresponding to the various scattering processes occurring in the system. 
The total state-to-state $(n_i,\ell_i) \to (n,\ell)$ scattering amplitude ${A}^S_{n,\ell}$ is split into a few terms according to the representation~(\ref{psi_sum}) of the total wave function $\Psi$,
\begin{equation}
{A}^S_{n,\ell} = [A_0^R]_{ n, \ell} + [A_1^R]^S_{n,\ell} + {\tilde{A}}^S_{n,\ell}.
\end{equation}
The term $[A_0^R]_{ n, \ell}$ corresponds to the function $\Psi^R_0$ and is calculated explicitly using the representation~(\ref{psi0_0}) of $\Psi^R_0$ for $r_1 > R$:
\begin{equation} \label{localAmpl_A0}
[A_0^R]_{ n, \ell} = \delta_{n, n_i} \delta_{\ell, \ell_i} 
({\cal A}^R_\ell + {\cal A}^C_\ell ).
\end{equation}
Here ${\cal A}^R_\ell$ is defined in Eq.~(\ref{amplA^R_l}), and the partial Coulomb scattering amplitude is given by
\[
{\cal A}^C_\ell = \frac{\exp{(2i\sigma_\ell)}-1}{2i}.
\]
The terms $[A_1^R]^S_{n,\ell}$ and ${\tilde{A}}^S_{n,\ell}$ correspond to the functions $\Psi^R_1$ and $\Phi^S=\Phi^S_0 + \Phi^S_1$, respectively.

The method used to calculate the amplitudes is derived from the asymptotic form of the scattered wave function at large distances~\cite{FadMerk},
\begin{eqnarray} \label{asympt}
\Phi^S(r_1,r_2,\theta) \sim
 \sum\limits_{n,\ell} {\tilde{A}}^S_{n,\ell} 
 \frac{1}{\sqrt{2}} \left(1+(-1)^S P_{12}\right)
 \frac{\varphi_{n,\ell}(r_2)} {r_2}
 \nonumber \\
 \times u^+_{\ell}(\eta_n, p_n r_1) 
Y_{\ell,0}(\theta,0) + B(r_1, r_2, \theta),
\end{eqnarray}
where the function $B(r_1, r_2, \theta)$ represents the three-body ionization term.
For large hyperradius $\rho=\sqrt{r_1^2+r_2^2}$, it decreases as $B(r_1, r_2, \theta) \sim \rho^{-1/2}$.
Projecting the representation~(\ref{asympt}) on the two body wave functions and taking into account the orthogonality of the two and three-body states, we obtain the local representation for the partial amplitudes ${\tilde {A}}^S_{n,\ell}$ for large $r_1$:
\begin{eqnarray} \label{localAmpl}
{\tilde{A}}^S_{n,\ell} \approx \sqrt{2}\frac{\Big( u^+_{\ell}(\eta_n, p_n r_1) \Big)^{-1}}{2\pi}
\int\limits_0^{\infty}dr_2 \int\limits_0^{\pi} \sin\theta d\theta
\nonumber \\
\times r_2 \, \varphi_{n,\ell}(r_2) \Phi^S(r_1,r_2,\theta) Y_{\ell, 0}(\theta,0)  .
\end{eqnarray}
The symmetrized term is neglected as the discrete state wave function decreases exponentially with $r_1$.
This representation is also used in order to calculate $[A_1^R]^S_{ n,\ell}$, where the function $ \Phi^S(r_1,r_2,\theta)$ is replaced with the function $\Psi_1^R (r_1,r_2,\theta)$.
In the calculations, we use the maximum value of $r_1$ available, i.e. $r_1=Q$.
The spin weighted cross section is then given in terms of the amplitudes by
\begin{equation} \label{crossec}
\sigma_{n \ell}^S = \frac{2S+1}{4} \frac{p_n}{p_i} |{A}^S_{n,\ell}|^2.
\end{equation}

With this approach, the solution of the scattering problem becomes a two-stage process.
First, the ECS method with $Q \ge R$ is applied to the driven Eqs.~(\ref{eq_psiR1},\ref{Phi_i_eq}) as discussed in~\cite{r:ecs3, ELY-helium}.
In this method, each of the spatial coordinates is replaced with the complex one $r \to s_\phi(r)$, where $\phi$ is the asymptotic rotation angle.
Any function $u(r)$ is then transformed as $(W^\phi u)(r) = \sqrt{|J(\phi)|} u[s_\phi(r)]$, where $J(\phi)$ is the Jacobian $J(r)=d s_\phi(r)/dr$.
The rotated Hamiltonian takes the form $H(\phi)= W^\phi  H  (W^\phi)^{-1}$.
The rotated Eqs.~(\ref{eq_psiR1},\ref{Phi_i_eq}) can be written as
\begin{equation} \label{driven_ECS}
(H(\phi) - E) (W^\phi F) = - W^\phi (\rm{R.H.S.}) .
\end{equation}
Both the solutions and the r.h.s. decrease at infinity, so these equations need to be solved with the boundary conditions equal to zero at infinity.

Furthermore, the scattering amplitudes are calculated from the non-rotated spatial part of the solutions, i.e. $r_1, r_2 \le Q$, with representations~(\ref{localAmpl_A0},\ref{localAmpl}).
This means that the asymptotic behavior~(\ref{asympt}) is used so $Q$ should be chosen large enough for this behavior to be valid.

\section{Numerical method and results} \label{results}

In order to demonstrate how our approach works numerically, the electron-H and electron-He${}^+$ systems have been chosen as examples.
The two systems are of fundamental importance in atomic physics so any newly developed approach is worth testing with them. 
The choice is naturally guided by the fact that these systems have been studied previously  (see~\cite{Henry1981, Bartlett2005} and references therein) using different methods and approximations.

In order to numerically solve Eq.~(\ref{driven_ECS}) for the scattering problem, the finite element method (FEM) has been employed~\cite{ELY-helium, LNCS}. 
In the calculations, we use a rectangular grid formed by the same one-dimensional grid in both coordinates $r_1$ and $r_2$.
For each coordinate, we use five finite elements at short distance [0--4]~a.u., and four elements of total length 40~a.u. for the discretization beyond the rotation point $Q$. 
The intermediate region [4--121]~a.u. is divided into elements with a length 3~a.u.
Only one element  was used for the angular variable $\theta$.
The polynomial degree on each element is chosen to be 7.
This implementation of FEM yielded a sparse matrix with a dimension up to 562176.
Details of our computational realization of the FEM can be found in~\cite{ELY-helium, LNCS}.

\begin{figure}[t]
    \centering \includegraphics[width=0.70\linewidth]{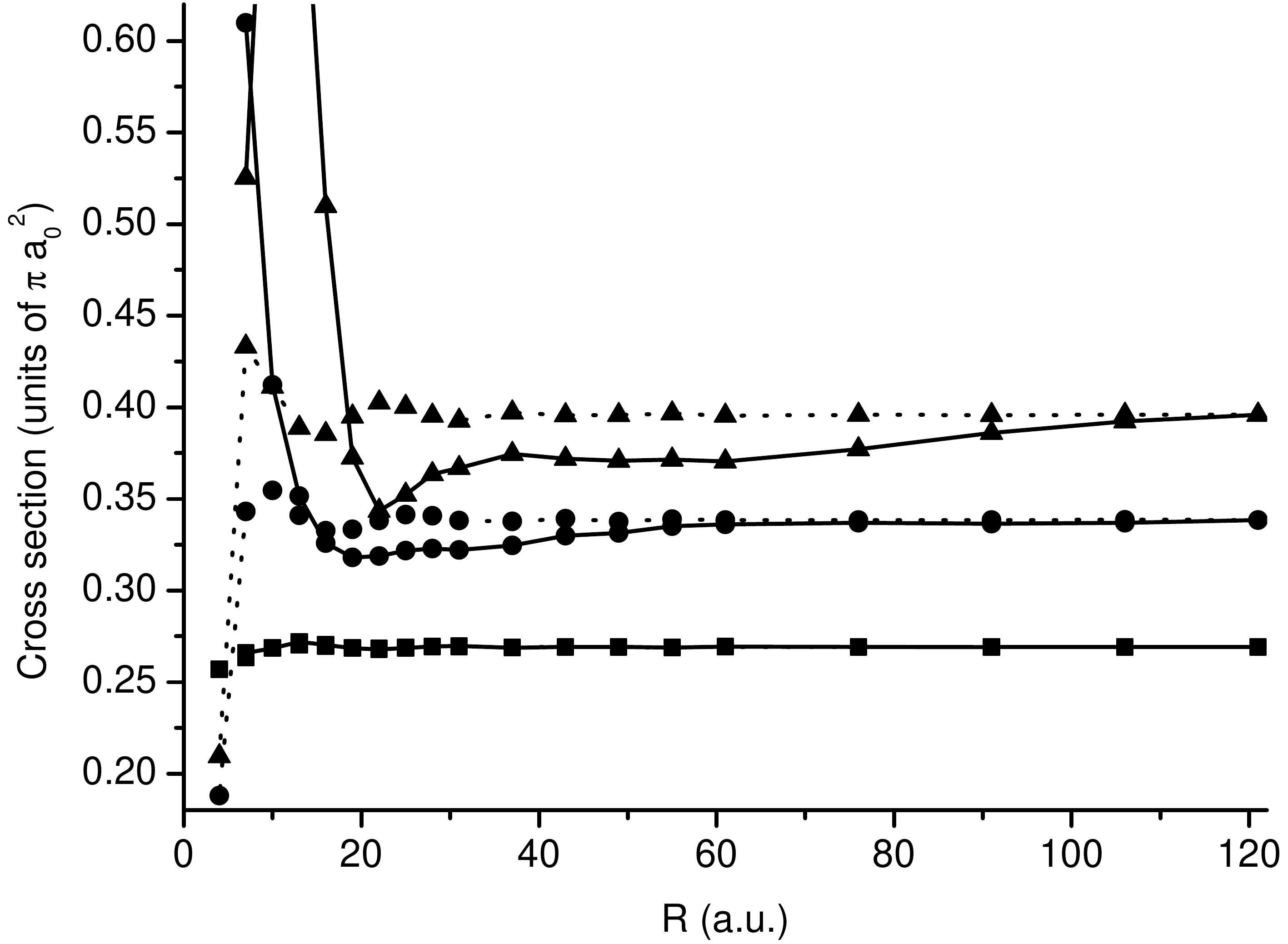}
\caption{The singlet (spin weight included) 1s$\to$1s (squares), 2s (circles), and 3s (triangles) cross sections for the e-H scattering as a function of the splitting radius $R$. 
    The solid lines correspond to the ECS radius $Q=R$, and the dotted lines do to $Q$=121~a.u.
    The values are multiplied by 20 and 80 for 2s and 3s cross sections, respectively.
    The energy $E$=17.6~eV. 
}
\label{fig1}
\end{figure}

\begin{figure}[t]
    \centering \includegraphics[width=0.70\linewidth]{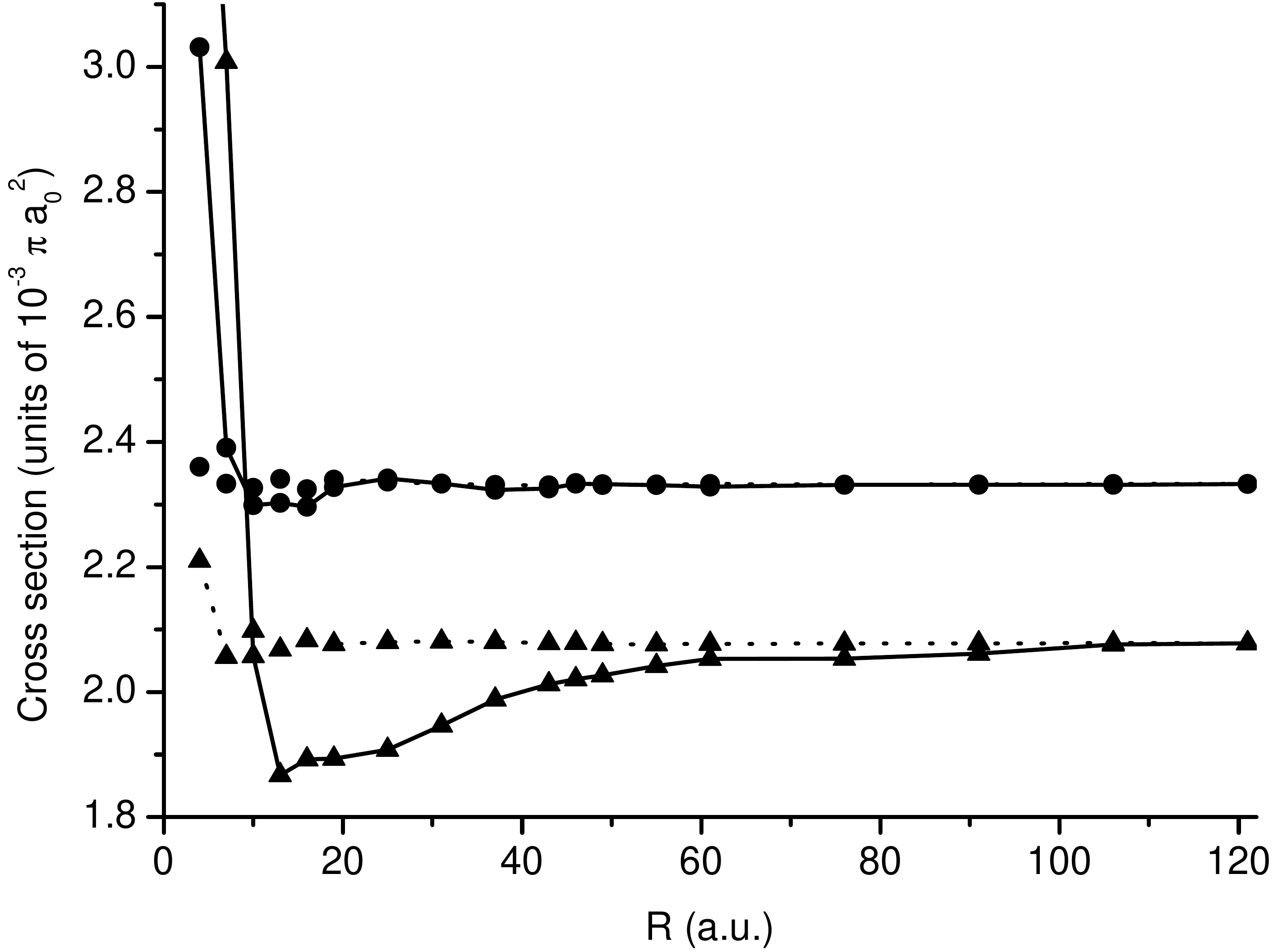} 
\caption{The singlet (spin weight included) 1s$\to$2s (circles) and 3s (triangles) excitation cross sections for the e-He${}^+$ scattering as a function of the splitting radius $R$.
    The solid lines correspond to the ECS radius $Q=R$, and the dotted lines do to $Q$=121~a.u.
    The values are multiplied by 3 for 3s cross sections.
    The energy $E$=17.6~eV. 
}
\label{fig2}
\end{figure}

As we need the solution of Eq.~(\ref{eq_psiR1}) with the ECS radius $Q' > R$, we first discuss the results for the calculations of the cross sections with different splitting $R$ and ECS $Q$ radii.
Our results for the singlet 1s$\to$ns cross sections for the e-H and e-He${}^+$ scattering are presented in Figs.~\ref{fig1} and~\ref{fig2}, respectively.
We do not consider the corrections $\Psi_1^R$ and $\Phi_1$ in this discussion.
We can see that, for a large ECS radius $Q$, the cross sections have high accuracy already for relatively small values of $R$.
This means that the splitting procedure itself does not introduce large inaccuracies in the amplitude calculations.
On the other hand, in the case $Q=R$, the cross sections are stabilized for large values of $R$ only.
As the amplitudes are calculated at the distance $Q$ with the asymptotic representation~(\ref{asympt}), it means that the main contribution to the inaccuracy in the amplitude originates from the inaccuracy in this asymptotic representation. 
This inaccuracy can be reduced with the use of the integral representation for the amplitudes when available.

The size of the excited two body Coulomb state grows as the square of the quantum number, implying that the higher excited state the larger distance is necessary to reach a converged result.
The size of a given helium ion state is smaller than that of the similar hydrogen state, implying that the cross sections for the He${}^+$ scattering stabilize for smaller distances.
It is worth noting that results for both the hydrogen scattering excluding the asymptotic Coulomb interaction and the He${}^+$ scattering including this interaction  converge equally well with respect to the splitting radius $R$.
This means that our splitting procedure completely takes into account the asymptotic Coulomb interaction.

\begin{figure}[t]
    \centering \includegraphics[width=0.7\linewidth]{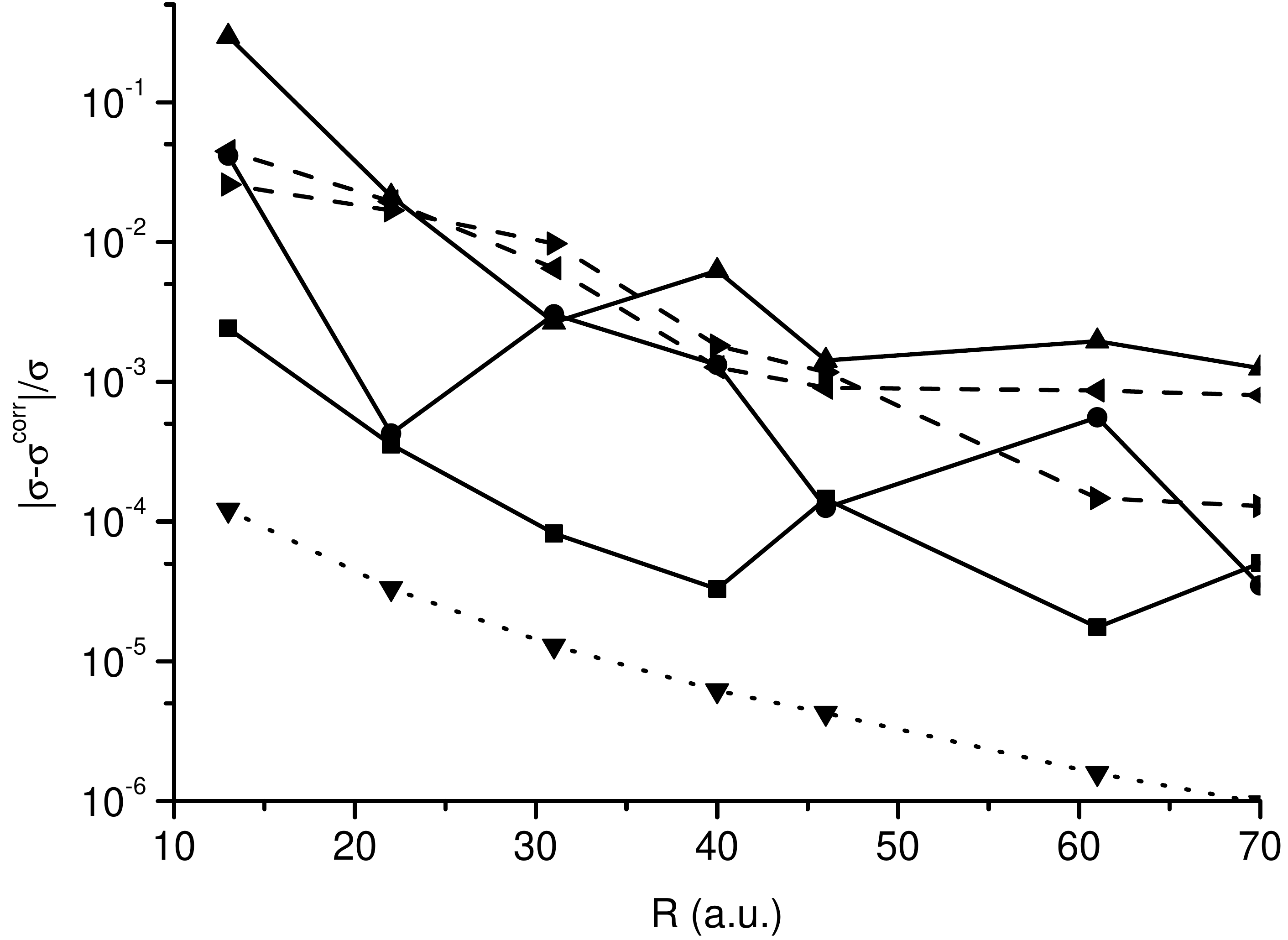} 
\caption{The relative difference $|\sigma-\sigma^{\rm corr}|/\sigma$ for the 1s$\to$1s (squares), 2s (circles), and 3s (triangles) cross sections for the e-H scattering as a function of the splitting radius $R$.
The ECS radius Q=121~a.u. 
The energies $E$=0.25, 0.46~a.u., and 0.647~a.u. correspond to the dotted, solid and dashed lines, respectively.
}
\label{fig3}
\end{figure}

\begin{figure}[t]
    \centering \includegraphics[width=0.7\linewidth]{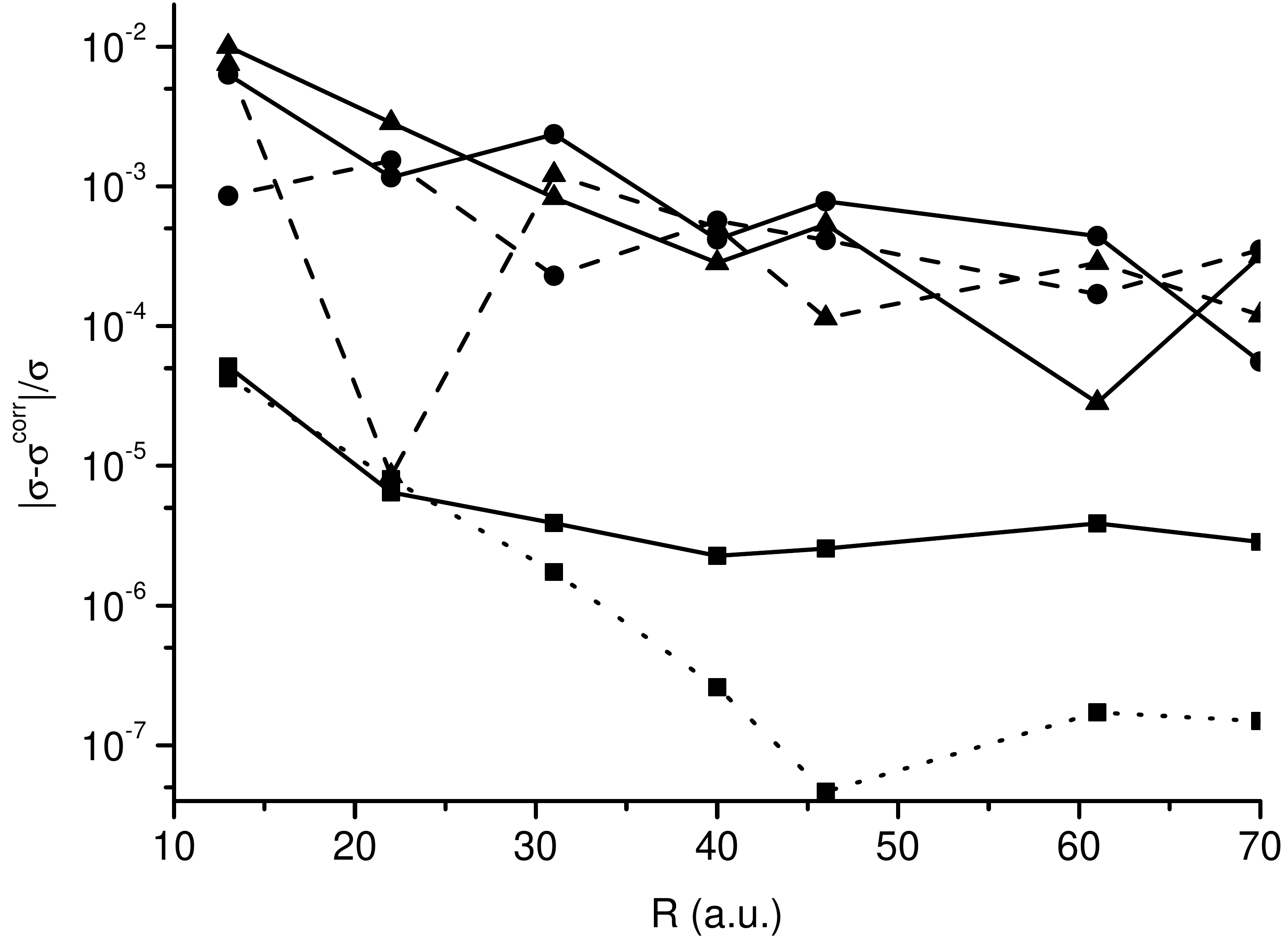} 
\caption{The relative difference $|\sigma-\sigma^{\rm corr}|/\sigma$ for the e-He${}^+$ scattering as a function of the splitting radius $R$.
The notations are the same as in figure~\ref{fig3}.
The energies $E$=1.0, 1.84~a.u., and 2.147~a.u. correspond to dotted, solid and dashed lines, respectively.
    }
    \label{fig4}
\end{figure}

The most important results for the justification of our approach are presented in Figs.~\ref{fig3} and~\ref{fig4}.
Here we show how the cross sections are influenced by the corrections $\Psi_1^R$ and $\Phi_1$.
We compare the cross section $\sigma$ calculated with the terms $\Psi^R_0 + \Phi_0$ only to the corrected cross section $\sigma^{\rm corr}$ calculated with the full wave function $\Psi$ in Eq.~(\ref{psi_sum}) which includes all corrections.
The relative differences $|\sigma-\sigma^{\rm corr}|/\sigma$ are shown in Figs.~\ref{fig3} and~\ref{fig4} for the e-H and e-He${}^+$ scattering, respectively.
We present the corrections for three typical energy regimes: (1) at an energy with only one open channel, (2) for an energy in the vicinity of resonance states, and (3) at an energy above the ionisation threshold. The calculated corrections diminish reasonably fast with $R$.
When only the elastic channel is open, the corrections are quite small and decrease quite fast.
If a few channels are open, the corrections get bigger and do not behave regularly due to the interaction between the channels. 
The rate of decrease does not depend on the scattering energy. The corrections follow the same trends with respect to the excitation number and to the system.

\begin{figure}[t]
  \centering \includegraphics[width=0.7\linewidth]{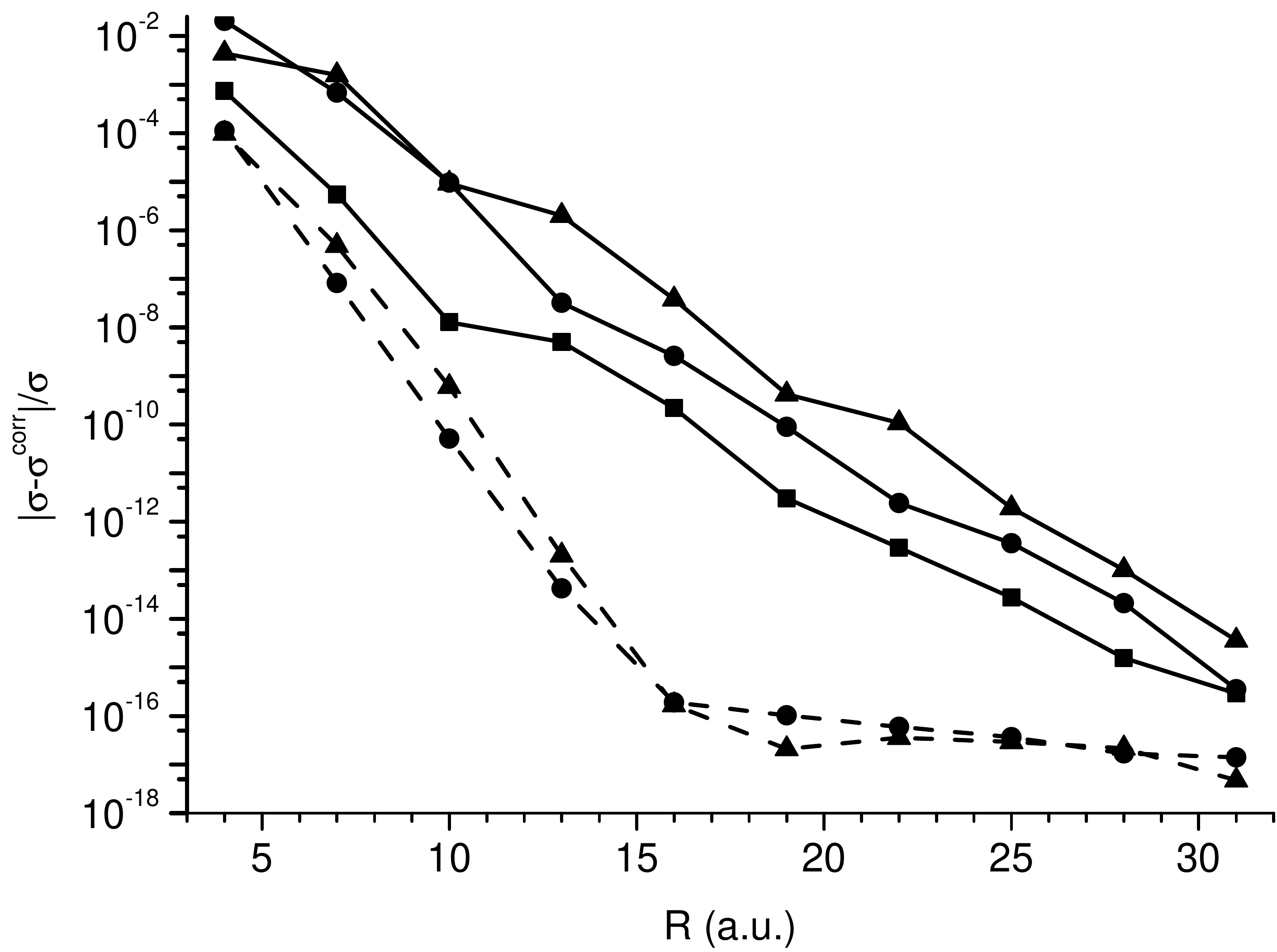} 
 \caption{The relative difference $|\sigma-\sigma^{\rm corr}|/\sigma$ for the 1s$\to$1s (squares), 2s (circles), and 3s (triangles) cross sections for the Temkin-Poet model as a function of the splitting radius $R$.
 The energy $E$=0.147~a.u. above the ionization threshold, the ECS radius Q=121~a.u.
 The solid and dashed lines correspond to the e-H and e-He${}^+$ scattering, respectively.}
\label{fig5}
\end{figure}

For comparison, we plot in Fig.~\ref{fig5} a few relative cross section differences for the Temkin-Poet model of the same systems~\cite{JPB-2015}.
One can see that the influence of the corrections is drastically smaller here and decreases very fast with the splitting radius $R$.
The reason for this is that the terms $\Psi_1^R$ and $\Phi_1$ decrease exponentially with $R$ for the Temkin-Poet model~\cite{JPB-2015} while they behave as inverse powers~(\ref{correction_estim}) for the full scattering problem.

\begin{figure}[t]
    \centering \includegraphics[width=0.7\linewidth]{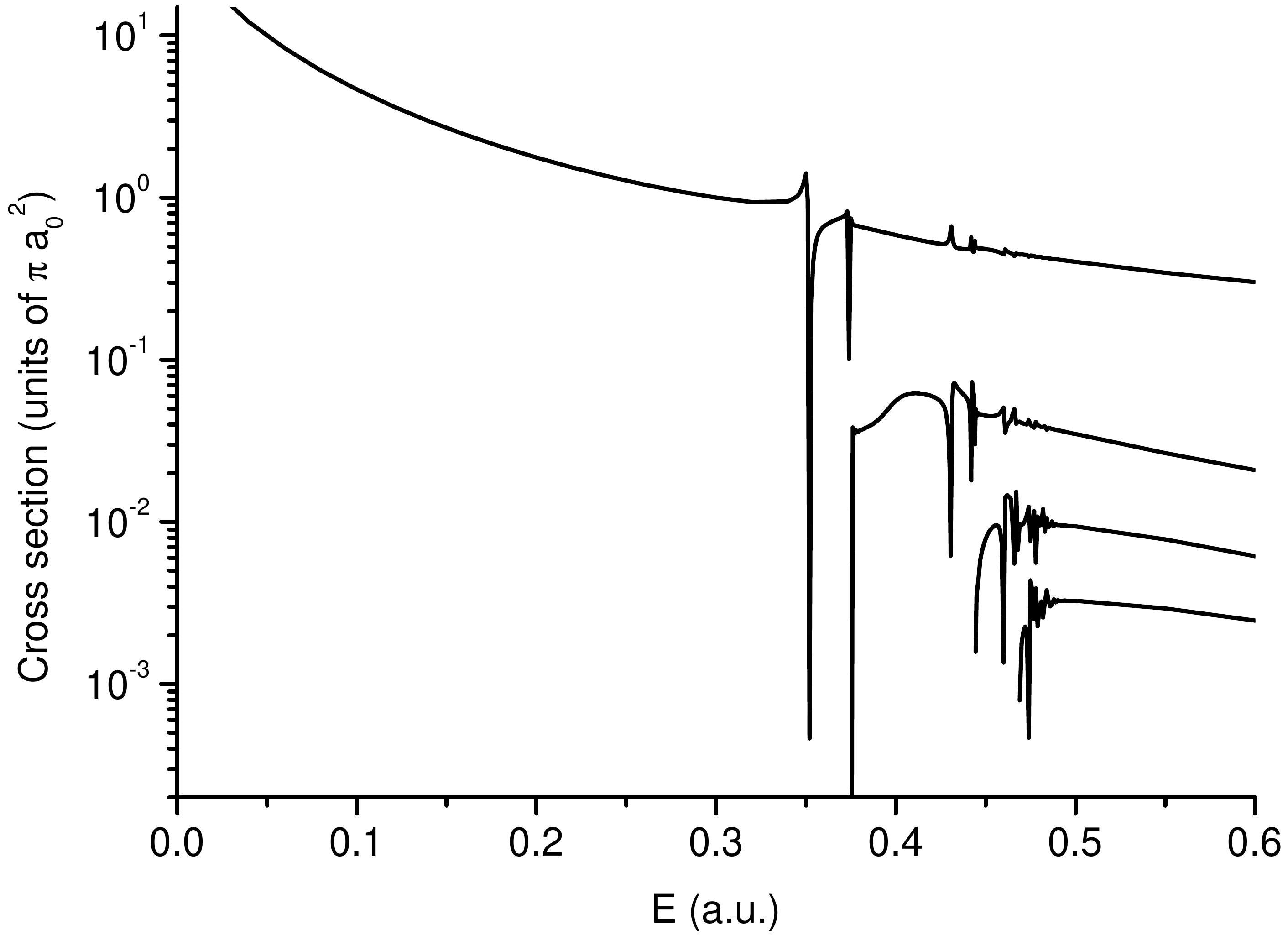} 
    \caption{The singlet (spin weight included) 1s$\to$1s, 2s, 3s, 4s cross sections (from above) for the e-H scattering as a function of the incident electron energy. The thresholds are 0, 0.375, 0.444 and 0.469~a.u., respectively.}
    \label{fig6}
\end{figure}

\begin{figure}[t]
    \centering \includegraphics[width=0.7\linewidth]{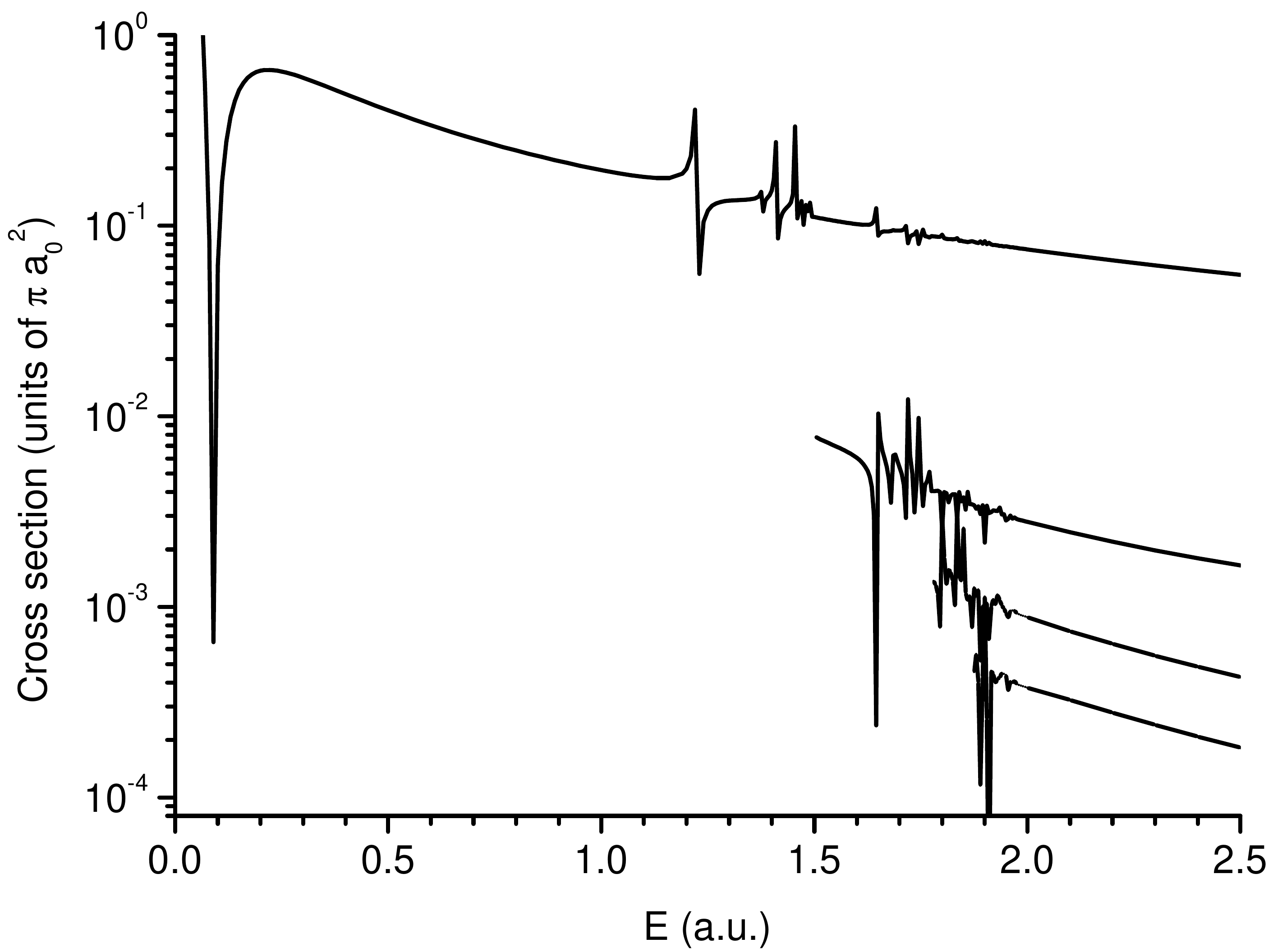} 
    \caption{The singlet (spin weight included) 1s$\to$1s, 2s, 3s, 4s cross sections (from above) for the e-He${}^+$ scattering as a function of the incident electron energy. The thresholds are 0, 1.5, 1.778 and 1.875~a.u., respectively.}
    \label{fig7}
\end{figure}

Our results for the singlet 1s$\to$ns cross sections for the e-H and e-He${}^+$ scattering are presented in Figs.~\ref{fig6} and~\ref{fig7}.
As we have seen in Figs.~\ref{fig3} and~\ref{fig4}, the relative inaccuracy for the cross section is below 10${}^{-3}$.
Therefore, the correction terms can be safely neglected. 
One can also see in Figs.~\ref{fig1}, \ref{fig2} that the best accuracy is achieved when the splitting and rotation radii are equal. 
Thus the results in Figs.~\ref{fig6}, \ref{fig7} are calculated with the values $R=Q=121$~a.u.
The cross sections have rich resonance structure, especially the cross section in the e-He${}^+$  scattering where the asymptotic Coulomb interaction is present.
Due to this interaction the cross section also exhibits oscillating behaviour at small energies.
The calculated values were compared to other results~\cite{Bartlett2005, PRA-1992, PRA-1993}, and to the accurate data for electron-hydrogen elastic scattering in the vicinity of resonance states~\cite{JPhB29_L59_1996}.
The relative difference is found to be less than 10${}^{-3}$.

\section{Conclusions} \label{Conclusions}
Here we present a formalism for the application of the potential splitting method to solve a driven three-body zero total angular momentum Schr{\"o}dinger equation, which includes the long-range Coulomb interaction, and realised using the exterior complex scaling method. 
This has been applied to the e-H and e-He${}^+$ collision systems.

The total wave function is split into four components which pairwise describe the incoming and outgoing waves of the scattering process. 
The scattering amplitude is likewise split into four components.
The theory as it is derived here is complete, in the sense that no components are neglected.  
The contributions are analysed with respect to both the splitting $R$ and exterior scaling $Q$ radii.

We have  numerically demonstrated  that terms corresponding to non-factorisable part $\Psi^R_1$ of the distorted incoming wave~(\ref{Psi^R_10}) decrease with increase in $R$.
While these terms do not vanish as fast as for the Temkin-Poet model, they still can be neglected for moderate values of $R$ and $Q$.
When comparing our numerical results with previously published results~\cite{Bartlett2005, PRA-1992, PRA-1993, JPhB29_L59_1996}  we find that the relative differences are less than 10${}^{-3}$.
We can thus state that our potential splitting method allows us to obtain numerically exact solutions for the three-body scattering problem with a Coulomb interaction.
Generalization of the present total zero angular momentum formalism to a full total angular momentum method is now under development as is the subsequent extension to include several reaction channels.

\begin{acknowledgments}
E.Y. and S.L. Y. are grateful for  support given by the Russian Foundation for Basic Research grant No. 14-02-00326 and by St. Petersburg State University within the project No. 11.38.241.2015  while N.E acknowledges a grant from the Carl Trygger foundation.
\end{acknowledgments}


\begin{thebibliography}{99}

\bibitem{FadMerk}
L. D. Faddeev and S. P. Merkuriev, \textit{Quantum Scattering Theory for Several Particle Systems} (Dordrecht: Kluwer, 1993).

\bibitem{PhysRep.520.135}
I. Bray, D. V. Fursa, A. S. Kadyrov, A. T. Stelbovics, A. S. Kheifets, and A. M. Mukhamedzhanov, Phys. Rep. \textbf{520}, 135 (2012).


\bibitem{CS} E.~Balslev and J.~M.~Combes, Commun. Math. Phys. \textbf{22}, 280 (1971).
\bibitem{Rescigno1997} T.~N.~Rescigno, M.~Baertschy, D.~Byrum, and C.~W.~McCurdy, Phys. Rev. A \textbf{55}, 4253 (1997).


\bibitem{r:ecs1} 
B. Simon, Physics Letters \textbf{71A}, 211 (1979).
\bibitem{r:ecs2} 
P.~D.~Hislop and I.~M.~Sigal, \textit{Introduction to Spectral Theory: With Applications to Schr\"odinger Operators. Applied Mathermatical Science, Vol. 113,} Springer Verlag, New York, 1996.

\bibitem{Gasaneo2012}
G. Gasaneo, L. U. Ancarani, and D. M. Mitnik, Eur. Phys. J. D \textbf{66} 91 (2012)
\bibitem{Zaytsev2016}
A. S. Zaytsev, L. U. Ancarani, and S. A. Zaytsev, Eur. Phys. J. Plus \textbf{131} 48 (2016).


\bibitem{PRA-2011} 
M.~V.~Volkov, S.~L.~Yakovlev, E.~A.~Yarevsky, and N.~Elander, Phys. Rev. A \textbf{83}, 032722 (2011).
\bibitem{EPL-2015} 
M.~V.~Volkov, E.~A.~Yarevsky, and S.~L.~Yakovlev, Europhysics Letters \textbf{110}, 30006 (2015).
\bibitem{JPB-2015} 
E.~Yarevsky, S.~L.~Yakovlev, {\AA}.~Larson, and N.~Elander, J. Phys. B. \textbf{48}, 115002 (2015).
\bibitem{r:eyrescode} 
N.~Elander, M.~V.~Volkov, {\AA}.~Larson, M. Stenrup, J. Z. Mezei, E.~Yarevsky, and S.~L.~Yakovlev, Few-Body Syst. \textbf{45}, 197 (2009).

\bibitem{r:ecs3}
T.~Alferova, S.~Andersson, N.~Elander, S.~Levin and E.~A.~Yarevsky, Advances in Quantum Chemistry, \textbf{40}, 323 (2001).

\bibitem{ourJPhysA2010}
S.~L.~Yakovlev, M.~V.~Volkov, E.~Yarevsky, and N.~Elander, J. Phys. A: Math. Theor. \textbf{43}, 245302 (2010).

\bibitem{AbrSt}
M.~Abramowitz and I.~Stegun (Eds.), \textit{Handbook of Mathematical Functions}, Dover, New York, (1986).


\bibitem{CBR2004}
C. W. McCurdy, M. Baertschy, and T. N. Rescigno, J. Phys. B: At. Mol. Opt. Phys. \textbf{37}, R137 (2004).


\bibitem{ELY-helium} 
N.~Elander, S.~Levin, and E.~Yarevsky, Phys. Rev. A \textbf{67}, 062508 (2003).

\bibitem{Henry1981} 
R.~J.~W.~Henry, Phys. Rep. \textbf{68}, 1 (1981).
\bibitem{Bartlett2005} 
P.~L.~Bartlett, \textit{Complete numerical solution of electron-hydrogen collisions} (PhD Thesis, Murdoch University, 2005).
\bibitem{LNCS} 
E.~Yarevsky, \textit{LNCS 7125: Mathematical Modeling and Computational Science} (Springer, 2012) 290.

\bibitem{PRA-1992} 
J.~Botero, J.~Shertzer, Phys. Rev. A \textbf{46}, R1155 (1992).
\bibitem{PRA-1993} 
Y.~D.~Wang, J.~Callaway, Phys. Rev. A \textbf{48}, 2058 (1993).

\bibitem{JPhB29_L59_1996} 
Y.~D.~Wang, W.~C.~Fon and C.~D.~Lin, J. Phys. B. \textbf{29}, L59 (1996).

\end{thebibliography}
\end{document}